\documentstyle[12pt]{article}

\parskip=\medskipamount
\begin{document}
\baselineskip=20pt

\begin{center}
{ \large \bf NONLINEAR DYNAMICS IN THE BINARY DNA/RNA CODING PROBLEM }
\footnote { Presented at the 7$^{th}$ Int. Summer School on Biophysics,
Rovinj, Sept.14.-25. 2000.}\\ 
\vspace{1.5cm}
{ \normalsize \bf M. Martinis, \footnote { e-mail address: martinis@thphys.irb.h
r}}\\
Department of Physics, Theory Division, \\
Rudjer Bo\v skovi\' c Institute, P.O.B. 1016, 10001 Zagreb,CROATIA \\
\end{center}

\noindent
 { \large \bf Abstract.} The presence of low dimensional chaos in the protein secondary structures, using the binary coded 
$\alpha $-helices and $\beta $-sheet motifs, has been investigated. In order to analyse symbolic 
DNA/RNA sequences the  assignment, based on the purine/pyrimidine and strong/weak H - bond classification sheme, 
of binary numbers to the nucleotides was made.  We have found that this assignment leads to different averaged 
fractal dimensions for $\alpha $-helices and $\beta $-sheets. When embedded in the two dimensional 
phase space the binary string structures show the attractor type behavior which depends on the value of the 
time lag parameter. 

\vspace{0.5cm}
\baselineskip=20pt

\vspace{1cm}

\newpage
\baselineskip=24pt
\section{Introduction}

The concepts of chaos, nonlinearity and  fractal can be  applied to a number of properties of proteins. 
Recently, there has been a rapid accumulation of new information in this area [1,2]. 
However, since proteins are finite systems, although highly inhomogeneous in their structure,  
their fractal description can only be true  in an average or statistical sense.  Even if a fractal  description 
is applicable, the question, 'What new insight does it provide?'  is sometimes necessary to 
answer.  For proteins this answer is not always  clear and straightforward. Nevertheless, deterministic chaos if
detected offers a striking explanation for irregular behaviour and anomalies in the protein structure 
which can not be attributed to the randomness alone.  

A protein consists of a polypeptide chain that is made up of residues of amino acids (20 commonly 
occuring that differ in their side-chains) linked together by peptide bonds. What distinguishes 
different proteins, other than the number of amino acids, is the sequence of amino acids, referred 
to as the primary structure of the protein, in the 
polypeptide chain. From the analysis of many protein structures determined by X-ray crystalography, 
it is found that the folding of portions of the polypeptide chain often has certain regularities, 
referred to as elements of secondary structure. The most important secondary structural elements 
are the $\alpha $-helices and $\beta $-pleated sheets. The $\alpha $-helix is a compact rod like 
structure, whereas the $\beta $- sheet is an extended structure. Since globular proteins have 
finite size , the secondary structural elements ( $\alpha $-helix , $\beta $-sheet) are also limited in lenght. 
Hence the validity of a fractal description as applied to these structures must be taken with care.

In this paper we shall, as one of the tests for nonlinearity, calculate the fractal dimensions for 
different secondary structure elements of  proteins  ($\alpha $-helices and $\beta $-sheets). 
We shall show that  fractal dimensions can be used to discriminate an 
$\alpha $-helix from a $\beta $-sheet. We shall also show how the two-dimensional embedding of the structure can 
produce an attractor type behavior. The paper is organized in the following way. In Section 1. we introduce the 
basic mathematical tools  necessary for nonlinear analysis of  finite protein sequences treated 
as finite time series. In Section 2. we apply 
this formalism to DNA/RNA and protein sequences and study a particular aspect of it by finding
the fractal dimensions of $\alpha $-helices and $\beta $-sheets using  the Hurst exponent approach [4]  which
measures the smoothness of a sequence.

\section{Nonlinear analysis of finite sequences }

The most direct link between chaos theory and the real world is the analysis of measured signals/ time series
in terms of nonlinear dynamics. In this sense, each nucleotide sequence is 
represented by a string of  suitably chosen numbers u(n), n = 1,...,N which give a novel 
representation of a sequence that can be further studied using techniques
derived from nonlinear analysis [3]. Next step is to detect and quantify possibly complicated structures in a sequence
which are real and not just random fluctuations. The most convincing argument for the presence of some pattern 
in the sequence is if it can be used to give an improved prediction of a next term in the sequence. 
A simple prediction algorithms always assume
some deterministic structure in the sequence which can be either linear  based on the 
assumption that the next observation is given approximatelly by a linear combinations of the previous ones plus noise,
$\xi $, i.e.,
\begin{equation}
u(n+1) = a_{0} +  \sum_{k=1}^{m}a_{k}u(n-m+k) + \xi(n), 
\end{equation}
or nonlinear
\begin{equation}
u(n+1) = F[u(n) + \xi(n)]
\end{equation}  
where the mapping F is usually an unknown nonlinear function. In that case  some additional assumption about the   
properties of F (e.g., continuity) and $\xi(n)$ (e.g., white noise) should be made. 
One test for nonlinearity is to compare the linear and 
nonlinear models to see which more accurately predict the next term in the sequence [5].
However, in cases where the considered sequences are not manifestly deterministic and are in addition finite and noisy 
like in biology, physiology and 
medicine the usefulness of chaos theory becomes  controversal.  

It is clear that, a priori, we cannot assume existence of deterministic chaos for any real data that 
are finite and noisy. The crutial step in resolving this problem is to look for signatures of determinism 
which can tell us how far we are from the linear case and how close we are to the nonlinear 
deterministic case. The usual estimators of nonlinearity are:
\begin{itemize}
\item {\bf nonstationarity}
\item {\bf fractal dimensions}
\item {\bf phase space embeddings}
\item {\bf Lyapunov exponents.}
\end{itemize}

\subsection{Nonstationarity}

A scientific measurement of any kind is only useful if it is reproducible, at least in principle.
Reproducibily is closely connected with the problem of stationarity  of sequence measurements.
If the process under observation is a probabilistic one, it will be characterised by probability
distributions. Stationarity means that they  are independent of time translations.
 
Almost all methods of time series analysis,linear or nonlinear, require some kind of stationarity.
For finite sequences the stationarity is not well defined and dependents on the lenght 
of the sequence.  
Most of the tests for stationarity are based on the estimate of a certain parameter using different 
segments of a sequence. If the observed variations are outside the expected statistical fluctuations
( standard error), the sequence is considered as nonstationary. In many applications of linear statistics,
weak stationarity is assumed which means that the mean and the variance of a sequence are constant.
Strict stationarity in addition also requires that the sequence is normally distributed.  
Therefore, the obvious test for nonstationarity is to consider the length dependece of the mean and the
variance of  different segments of a sequence:
\begin{eqnarray}
\bar{u}_i(l) & = & \frac{1}{l}\sum_{k=1}^{l}u(il - l + k) \\
\sigma_i^{2}(l) & = & \frac{1}{l(l+1)}\sum_{k=1}^{l} (u(il -l + k) - \bar{u}_{i}(l))^{2}
\end{eqnarray}
where the index i denotes the i-th segment of the sequence u(n) having the length l. For these two quantities
the ordering of elements inside the segment is irrelevant and thus they cannot give any information 
about the evolutin of a system. Such information can be obtained from the sample autocorrelation function
\begin{equation}
C(n) = \frac{1}{N-n}\sum_{k=1}^{N-n}(u(k) - \bar{u})(u(n + k) - \bar{u})
\end{equation}
which measures, at fixed  lag n, how the points u(k) and u(n + k) are distributed in the plane.
 
Autocorrelation can be detected visually by examining a regression line scatterplot. Distances from the
regression line to the actual values represent error. When no autocorrelation exists, dots representing actual values
will be randomly scattered around the full length of the series. 
The value of the autocorrelation function at lag n=0  is the variance $\sigma ^{2}$ of the sequence.

The autocorrelation function C(n) may be used to detect deterministic components masked in a random
background because deterministic data usually have  long range correlations in which C(n)$\sim  n^{-\beta }$
for large n, while random data have  short-range correlations and C(n) decays exponentialy for large n. 
Observation of long-range correlations in a sequence may be a first sign for underlying nonlinear dynamics.

Another useful characteristic of nonstationarity is the  behaviour of the low-frequency  part of the 
power spectrum which is defined as the Fourier transform of the autocorrelation function C(n).
An equivalent definition of the power spectrum is the squared modulus of the Fourier transform of a 
sequence u(n):
\begin{equation}
P(f) = |\frac{1}{\sqrt{N}}\sum_{n = 1}^{N} u(n) e^{2\pi if n}|^{2}
\end{equation}
where the frequencies in physical units are f$\equiv $  f$_k$ = k/N, k = -N/2,...,N/2. The relation 
between the power spectrum P(f) and the autocorrelation function C(n) is known as the Wiener-Khinchin 
theorem. The sequencies with the power spectra  at  low frequences proportional to $|f|^{-\alpha }$  with
positive $\alpha $  are considered nonstationary. This trend usually characterizes the most biological 
signals. White noise  is defined by a flat spectrum; it has  a 
spectral exponent $\alpha $= 0. 
When white noise is summed up over n, the result is a random walk process  with a spectral 
exponent $\alpha $= 2, known as a brown noise. 
The intermediate case with $\alpha $= 1, known as a pink noise, is encountered in
a wide variety of physical, chemical, and biological systems. It implies that all lenght-scales 
in a sequence, over which 1/f behaviour holds, are equally important. A power spectrum with  $\alpha >$ 2  is
known as a black noise; it governs the phenomena connected with natural and unnatural catastrophes
which often come in clusters.
  
For fractal sequences to be discussed below there is a simple relation between the exponent $\alpha $
and the fractal dimension D of the sequence  given by $\alpha $ = 5 - 2D .

\subsection{Fractal Dimensions}

The recent realization that nonlinear dynamical systems can display deterministic chaos has had a strong
influence  on the researchers to interpret complex physiological data in terms of chaos. A number of
techniques based on concepts from nonlinear dynamics have been proposed [6]. The fractal dimension [7] as 
one of them gives a statistical measure of the geometry of the clouds of points. In deterministic chaotic systems
the dimension is not always an integer number but fractional. Therefore, fractals are disordered systems
whose disorder can be described in terms of noninteger dimensions. The focal point of fractal geometry
is the repetition of disorder at all length scales (scale symmetry), or simply,  
the repetition of structure within structure. There are various ways and levels of precision of defining 
and calculating the fractal dimensions. We shall present here the calculation of the fractal dimension of
an irregular sequence by means of the Hurst's rescaled range analysis exponent [4].  

For a given sequence u(n), n = 1,...,N the rescaled range is defined as the range R(n) of the sum of the 
deviations of data from the mean divided by the sample standard deviation S(n). The relevant quantities
are defined in the following way. 
The sample mean over the length n is 
\begin{equation}
\bar{u}(n) = \frac{1}{n}\sum_{k=1}^{n} u(k),
\end{equation}
it is used to construct the accumulated departure X(l,n) of u(k) from the mean $\bar{u}(n)$,
\begin{equation}
X(l,n) = \sum_{k=1}^{l}[u(k) - \bar{u}(n)].
\end{equation}
The range R(n) is obtained as the difference between the maximum and the minimum of X(l,n) for fixed n,
\begin{equation}  
R(n) = \max_{l}X(l,n) - \min_{l}X(l,n).
\end{equation}
The standard deviation S(n) is estimated from
\begin{equation}
S(n) = \left(\frac{1}{n}\sum_{k=1}^{n}[u(k) - \bar{u}(n)]^{2}\right)^{\frac{1}{2}}.
\end{equation}
The rescaled range is then the dimensionless ratio R/S. The Hurst exponent H is defined in the
range of n where the Hurst law R/S $\sim $ n$^{H}$ is valid. In general, H is not a constant but 
a n-dependent quantity which can be  estimated using the log(R/S) versus log(n) plot. For these 
reasons the R/S analysis is considered as robust. The Hurst 
exponent is used to estimate the smoothness of an irregular  sequence. Thus,\\
if H = 0.5, the behaviour of the sequence is similar to a random walk;\\
if H$ <$ 0.5, the sequence is very spiky and is dominated by short-range fluctuations;\\
if H$ >$ 0.5, the sequence is less spiky and is dominated by long-range fluctuation.\\
There is a simple relation between H and the fractal dimension D:
\begin{equation}
D = 2 - H.
\end{equation}
H is  related to the "1/f" spectral slope $\alpha $ = 2H + 1 and can  also be estimaded by means of the
Detrendet Fluctuation Analysis (DFA) [8] used for quantifying the correlation properties in a nonstationary
sequence. DFA is based on the computation of a scaling exponent "d" by means of a modified root mean 
square analysis of a random walk: d = H + 1.
   
\subsection{Phase space embeddings}

When analysing a finite sequence u(n), n = 1,...,N, we  will almost always have only incomplete 
information about the underlying dynamical system which may live in a high-dimensional state space.
Therefore, in order to construct a dynamical model for a given sequence of data it is necessary first
to reconstruct a state space. One technique to do so is, for example,  the phase plane embedding or phase
portrait of a sequence, which involves plotting u(n + $\tau $) vs. u(n) with time lag $\tau $ , 
giving a trajectory in the two-dimensional phase space. Embeddings of a sequence in higher dimensional
phase space rest on the assumption that a genuine trajectory {\bf x}(t)      
of a dynamical system is a state vector in a d-dimensional space. The  delay embedding theorem 
by Takens [9] states that a $\tau $-delayed reconstructed state vector 
\begin{equation}
{\bf u}(n) = (u(n), u(n-\tau ), ..., u(n - E\tau  + \tau ))
\end{equation}
in E-dimensional embedding phase space will, in the absence of noise, approximate dynamical system
provided E $\ge $ 2d + 1; in practise  E $>$ 2d$_{Box}$, where d$_{Box}$ is the box counting fractal
dimension of a sequence [10]. For a deterministic dynamical system a phase-space description is an 
optimal way of studying its dynamical and geometrical properties. The problem of finding a good 
embedding is to estimate the embedding dimension E and to find the correct time lag $\tau $ [3].
  
\subsection{Lyapunov exponents}

So far we have discussed predictive criteria for finding nonlinear dynamics in a sequence of data.
A parameter for diagnosing whether or not a dynamical system is chaotic are the Lyapunov exponents 
which measure the average local exponential growth rate of infinitesimal initial errors, that is, of 
neighboring trajectories in phase-space embeddings [11].

If {\bf u}(n$_{1}$) and {\bf u}(n$_{2}$)  are two points in state space with the Euclidean distance 
$\parallel {\bf u}(n_{1}) - {\bf u}(n_{2})\parallel$ = d$_{0} \ll 1$ then the distance after a
time $\delta $n between the two trajectories will be $\parallel {\bf u}(n_{1} + \delta n) - 
{\bf u}(n_{2} + \delta n)\parallel$ = d$_{\delta n}$. Then the Lyapunov exponent $\lambda $ is defined by 
\begin{equation}
d_{\delta n} \simeq d_{0} e^{\lambda \delta n},
\end{equation}
with d$_{\delta n} \ll 1$ and $\delta n \gg 1$. If a system is known to be deterministic, a positive 
Lyapunov exponent can be taken as a definition of a chaotic system. Critical issues in the reliability 
of the estimation of Lyapunov exponents are the influence of noise.

\section{Fractal dimensions of DNA sequences in all-$\alpha $ and all-$\beta $ proteins}

In this Section we report on the finding the  fractal dimensions of DNA sequences appearing in the 
secondary strucures of proteins [12]. We have analysed 12 $\alpha $-helices and 12 $\beta $-sheets
using the Hurst exponent approach which measures the smoothnes of a given  sequence. Since the Hurst 
exponent H is related to the fractal dimension D = 2 - H of the sequence, to its "1/f" spectral slope 
$\alpha $ = 2H + 1 and to the DFA slope d = H + 1 it is a suitable parameter for studying
correlation properties in DNA sequences.

A DNA sequence consists of four kinds of bases: A, C, G and T/U. The bases can be divided in classes:
purine (R = A, G), pyrimidine (Y = C, T/U), strong H-bonds (S = G, C) and weak H-bonds (W = A, T/U).
In order to analyse symbolic DNA sequences the assignment of numbers to the nucleotides A, C, G and T/U 
is necessary. Taking into account the classes we choose the following binary assignment:
\begin{eqnarray}
Y \to  0  & , &  S \to  1 \\
R \to  1  & , &  W \to  0.
\end{eqnarray} 
This assignment maps the simbolic nucleotide sequence on a sequence u(n), n = 1,..., N(length of the sequence)
where u(n) can take values 0 or 1 only. Similar approach, based on the purine--pyrimidine classification sheme,
has been proposed by Peng et al.,[13]. Our analysis has shown that $\alpha $-helices and $\beta $-sheets have
different fractal dimensions, D$_{\alpha } < $ D$_{\beta }$ with  D$_{\alpha } \approx $1.41.

We have also studied the behaviour of sequences in 2-dimensional phase-space embeddings by first  transforming 
the digital sequence u(n) into an analoge signal  by means of the scaled sinc(t) function:
\begin{eqnarray}
A(t) & = & \sum_{k=1}^{N}u(k) sinc(\frac{t}{2} - k), \\
sinc(t) & = & \frac{sin(\pi t)}{\pi t}.
\end{eqnarray}
The phase portrait A(t + 1) vs. A(t) shows the attractor type behaviour.
   
\section{Conclusion}

In this paper we have discussed several simple methods which are useful for  analysing and estimating 
complex structures in DNA sequences. We have argued using the Hurst exponent approach that fractal
dimensions of DNA sequences in all-$\alpha $ and all-$\beta $ proteins are different. Observation of
scale dependent fractal dimension ( breaks in the slopes of log--log plots of R/S ) may indicate
a transition from one pattern of variation at one scale to another pattern at a larger scale. There are
probably several contributing causes to these observed variations in D. Although the DNA sequences
exibit statistical characteristic which might be explained by nonlinear statistical methods, 
no definite sign of chaos has been evidenced yet. 
  
\newpage

{ \large \bf Acknowledgment }

This work was supported by the Ministry of Science of the Republic
of Croatia under Contract No.980104.


\newpage

\begin{thebibliography}{99}

\bibitem{1}
R. Elber, in The Fractal Approach to Heterogeneous Chemistry, ( D. Anvir, ed.) John Willey \& Sons, NY 1990;

\bibitem{2} 
Proceedings of Santa Fe Institute Studies in the Sciences of Complexity, Vol. XV, 
(A. S. Weigend and N. A. Gershenfeld, eds.) Addison-Wesley, Reading, MA 1993;

\bibitem{3}
H. Katz and T. Schreiber, Nonlinear Time Series Analysis, Cambridge University Press, Cambridge, UK 1997;

\bibitem{4}
J. Feder, Fractals, Plenum Press, NY 1989;

\bibitem{5}
M. Casdagli and A. S. Weigend, in Proc. of Santa Fe Institute Studies in the Sciences of Complexity, Vol. XV,
Addison-Wesley,Reading, MA 1993; 

\bibitem{6}
P. Grassberger et al., Intl. J. Bif. \& Chaos {\bf 1} (1991) 512;

\bibitem{7}
J. D. Farmer, Physica {\bf 4D} (1982) 366,\\
P. Grassberger and I. Procaccia, Physica {\bf 9D} (1983) 189,\\
P. Grassberger, Phys.Lett. {\bf A97} (1983) 227;

\bibitem{8}
C. K. Peng et al., Chaos {\bf 5} (1995) 82;

\bibitem{9}
N. H. Packard et al., Phys.Rev.Lett. {\bf 45} (1980) 712,\\
F. Takens, in Dynamical Systems and Turbulence, (D. Rand and L.-S. Young, eds.), Springer-Verlag, Berlin 1981,\\
D. Ruelle, Proc.Roy.Soc. London Ser. {\bf A427} (1990) 241;

\bibitem{10}
T. Sauer et al., J.Stat.Phys. {\bf 65} (1991) 579;

\bibitem{11}
A. Wolf et al., Physica {\bf 16D} (1985) 285;

\bibitem{12}
M. Martinis, N. \v Stambuk and A. Kne\v zevi\' c, Fractal Dimensions of DNA--sequences, in preparation; 

\bibitem{13}
C. K. Peng et al., Nature {\bf 356} (1992) 168.
 
\end{thebibliography}
\end{document}